# FinTech E-Commerce Payment Application User Experience Analysis during COVID-19 Pandemic

**Leon A. Abdillah[1]**

[1]Information Systems Study Program, Faculty of Computer Sciences,
Universitas Bina Darma, Indonesia
Email: [1]leon.abdillah@yahoo.com

## Abstract

Application of information technology in the era of big data and cloud computing has led to the trend of electronic payments through financial technology, or FinTech. One of the most popular FinTech applications in Indonesia is Go-Pay in the Gojek start-up application. This research will analyze how the FinTech Go-Pay user experience both for transactions on Gojek and at merchants that collaborate with Gojek. User Experience (UX) is analyzed using the User Experience Questionnaire which consists of 6 (six) variables (Attractiveness, Perspicuity, Efficiency, Dependability, Stimulation, and Novelty). Total data collected amounted to 258. After analyzing the calculation results, the mean scores are obtained in the following order: Efficiency, Perspicuity, Stimulation, Attractiveness, Dependability, and Novelty. Then when compared with benchmark data the following sequence is obtained: Efficiency, Perspicuity, Stimulation, Attractiveness, Dependability, and Novelty. Overall the Go-Pay service is efficient and perspicuity, but the Go-Pay service needs to improve its novelty. This article provides additional knowledge or novelty contributions, especially for user experience analysis using FinTech applications.

**Keywords**: Cloud Computing, COVID-19, FinTech, Go-Pay, Online Transportation

## 1. INTRODUCTION

Progress in the development of information technology (IT) has entered a new era and changed many behaviors and habits. With IT, the transfer and exchange of data and information that were previously done manually can now be done online and connected regardless of time and distance. The internet allows sending data from source to destination in various forms. The internet has now become the global communication standard in all aspects of life, including in e-commerce and transportation sector, such as online ride sharing [1]. Adjusting information technology-based to human behavior [2] is the primary deal of Human Aspects, Ergonomics, Interaction of Human and Computer and Usability approach. the deliberative result of the user experience remains notable at the post-usage stage [3] for both web and mobile based applications.

The internet has also helped human activities when the corona virus outbreak struck in late 2019 and early 2020. The corona virus was originally called the "2019-novel coronavirus (2019-nCoV)", then by WHO officially on February 11, 2020 it was named "COVID-19" [4] caused by SARS-CoV-2 [5]. The COVID-19 outbreak has not only overwhelmed the country of China (as the beginning of the plague) [6], but even a number of developed countries were overwhelmed by it. Italy is the worst European country to be hit by COVID-19. The Kingdom of Saudi Arabia even limits the annual regular visits that have been held for thousands of years, namely the "Umrah". In fact, many countries are implementing the "LockDown" policy to limit the spread of COVID-19. But with the internet and also smartphones, as one of ubiquitous computing, has changed the trend of communication and transactions. Almost all applications that previously operated on a personal computer or laptop can be run from a smartphone alone. Even today the smartphone has shifted to the basic needs of every human being in urban areas especially. During the bleak period of COVID-19, many people's activities were carried out through the internet, laptops, and smartphones that were done from home (Work from Home, WFH).

Until early 2020, internet and smartphone users around the world are estimated to reach fantastic numbers. In 2020, it is predicted that the Pew Research Center will be dominated by the use of the internet on mobile devices [7]. Furthermore, smartphone [8] have been transformed into individual assistants, enjoyment gadgets, information portals. Indonesia itself is a country with high internets and smartphone penetration. The use of smartphones and personal computers in Indonesia is more related to the capacity for handling information in everyday life [9]. Indonesian internet users rank fourth in the world, third in Asia, and first in ASEAN.

Activities of internet users in surfing the web and social media have formed an ecosystem called the big data. This big data ecosystem requires a cloud-based platform [10]. Cloud computing [11], [12] is a powerful technology for large-scale and complex computing. The six existing start-up businesses (Table 1) all utilize big data and cloud computing technology.

Table 1. Top Indonesian startups

| No | StartUp | Category | Status |
|---|---|---|---|
| 1 | Gojek | Ride Hailing | Decacorn |
| 2 | TokoPedia | Market Place | Unicorn |
| 3 | Traveloka | Travel Site | Unicorn |
| 4 | BukaLapak | Market Place | Unicorn |
| 5 | OVO | Financial Technology | Unicorn |
| 6 | JD.ID | Market Place | Unicorn |

The existence of these new technologies has made a number of information technology-based businesses flourish in Indonesia. In fact, there is at least 1 (one) start-up that is classified as "decacorn" and 4 (four which is included in the unicorn category [13]. In 2020 one Market Place startup reaches the unicorn status. Table 1 shows 6 (six) top Indonesian startups that developed in the era of big data, cloud computing and ubiquitous computing.

Corona COVID-19 [6] virus sourced from Wuhan, China, has spread throughout the world. A number of countries have also begun implementing social distancing policies until lockdown. Transmission of the virus does not only occur due to proximity or contracting directly from the patient, but also indirectly contracting, for example through money intermediaries when payment transactions occur. So that minimizing the use of transactions with physical money should be done as much as possible. Transactions with physical money can be replaced by making transactions using virtual money in e-wallets and transacting with financial technology (FinTech).

The business sectors have been utilized IT as virtual media [14] both for sellers and buyers to create online transactions or virtual shopping. The new digital economic environment requires new payment transaction services as well. In other words, the transaction payment activities have also changed. Payments using physical money turn into virtual money made using financial technology (FinTech) applications. FinTech is the adoption of advances in information technology in the financial services industry [15]. National governments always play a critical role in fostering adoption of ICT-related indicators [16]. The adoption of information and communication technologies (ICT) fosters economic growth in Indonesia. FinTech appears to be significantly driving economic growth in their second year [17], and The FinTech industry also promises to reduce costs, as well as improve access to financial services [18]. In the FinTech era, all financial activities were digitized with an internet-based transaction mode and could be accessed by a website or smart device known as a smartphone. This research will continue the previous activity which analyzed the online transportation mobile application, Gojek [13]. Gojek itself as a technology-based business player from the beginning has introduced a digital payment system through the Go-Pay service on the Gojek application. This research will focus on payment transaction service, FinTech, which is provided by Gojek, Go-Pay. Through Go-Pay, Gojek has facilities for corporate financial services [19]. Go-Pay also considered successful in implementing financial inclusion and digital economy in Indonesia.

A number of studies have discussed Go-Pay as an alternative to digital payments. Previous research discusses how the influence of "trust" on users who will use online transactions [20]. They concluded that trust is affected by social influence and self-efficacy. Subsequent research discusses a descriptive analysis of Go-Pay users in Surabaya [21], which is dominated by women, age group Y generation, and high school education equivalent. Another research used technology Acceptance Model (TAM) to understand consumer intention to use Go-Pay [22]. And the last research discussed the process of adoption interest in using digital wallet of Go-Pay users in Central Jakarta [23].

None of the studies have raised a theme related to "user experience" when using financial technology services when conducting online transactions. That's why this study will analyze the user experience in using the FinTech application.

## 2. METHODS

The research approach applied in this study is a quantitative method that is combined with observations, interviews, and filling out the questionnaires online. The novelty used in this method includes research objects that have never been previously reviewed or analyzed with user experience.

### 2.1. Respondents

Respondents in this study were computer science students who belong to the Millennial and Z generations. The students have initial knowledge related to the development of technology based on big data, cloud computing, human-computer interaction (HCI), financial technology (FinTech), global positioning system (GPS), research methodology, smartphones, etc. The 21st century will indeed be dominated by social and business activities by the millennial generation and Z.

### 2.2. Research Object

The object of this research is the FinTech Go-Pay service that is available on the online transportation service, Gojek. Gojek is an ojek booking service through the Gojek application that can be downloaded on an Android Smartphone [24]. The modes of transportation between shuttles by using the media to call gadgets or smartphones online [25]. Go-Pay is used by Gojek customers to make virtual payments. Besides being commonly used for transactions on Gojek, Go-Pay can also be used for payment transactions at many merchants that work with Gojek.

Go-Pay can be imagined as an e-Wallet that can be filled in with the balance. With the balance in Go-Pay, users can use the balance the same as using physical money. Even in early 2020, Go-Pay has added PayLater service to its selected users. With the PayLater facility, selected Gojek customers can enjoy a variety of Gojek services with payments made at the end of the current month.

Payment for services in the Gojek application will be deducted automatically. Meanwhile, to make payments at merchants with Go-Pay, users can simply scan the barcode that is at the intended merchant. After a successful scan, then enter the nominal to be paid. If the transaction is successful, the Go-Pay balance will be reduced and the transaction slip will be printed from the machine at the merchant. With QR Code technology [26], it makes it easier for buyers to make payment transactions based on FinTech Go-Pay. In Indonesia, mobile commerce or m-commerce [27] is growing very rapidly and is now widely used by a number of customers. Gojek super application through mobile wallet, Go-Pay, has expanded financial access for millions of people in Indonesia.

### 2.3. User Experience Questionnaire (UEQ)

UEQ is a tool that can be used to measure User Experience (UX) for both web and mobile based applications. UEQ contains 6 (six) scales or UX variables, namely: 1) Attractiveness, 2) Efficiency, 3) Perspicuity, 4) Dependability, 5) Stimulation, and 6) Novelty [28]. Each scale is represented by an average of 4 (four) indicator items, except the Attractiveness variable contains 6 (six) items [29] (Table 2). This questionnaire is a measure that is easy to apply, reliable, and valid for the user experience [30].

UEQ is a semantic differential with a 7-point Likert scale for the answer. Each question item represents 2 (two) terms with opposite meanings [31]. The lowest score represents the most negative side, while the highest score represents the most positive value. The experience variables and indicators shown in Table 2.

Table 2. Experience variables and indicators

| No | Variables | Indicators |
|---|---|---|
| 1 | Attractiveness | 1) annoying/enjoyable, |
|   |   | 2) bad/good, |
|   |   | 3) unlikeable/pleasing, |
|   |   | 4) unpleasant/pleasant, |
|   |   | 5) unattractive/attractive, |
|   |   | 6) unfriendly/friendly. |
| 2 | Perspicuity | 1) not understandable/understandable, |
|   |   | 2) difficult to learn/easy to learn, |
|   |   | 3) complicated/easy, |
|   |   | 4) confusing/clear. |
| 3 | Efficiency | 1) slow/fast, |
|   |   | 2) inefficient/efficient, |
|   |   | 3) impractical/practical, |
|   |   | 4) cluttered/organized. |
| 4 | Dependability | 1) unpredictable/predictable, |
|   |   | 2) obstructive/supportive, |
|   |   | 3) not secure/secure, |
|   |   | 4) does not meet expectations/meet expectations |
| 5 | Stimulation | 1) inferior/valuable, |
|   |   | 2) boring/exiting, |
|   |   | 3) not interesting/interesting, |
|   |   | 4) demotivating/motivating |
| 6 | Novelty | 1) duil/creative, |
|   |   | 2) conventional/inventive, |
|   |   | 3) usual/leading edge, |
|   |   | 4) conservative/innovative |

### 2.4. Data Collection

Data is collected by utilizing social technology (SosTech) facilities from Google, Google Forms. After the questionnaire was created using Google Forms, it was then distributed through a number of Facebook Groups, WhatsApp Group, etc. Google Forms has the facilities needed by creating standard questionnaires including UEQ requirements. An example of how UEQ questions appear on Google Forms, as shown in Figure 1.

Figure 1. Example display questions on google form

### 2.5. Data Analysis
Google Forms [32] has the ability to display collected data in the form of worksheets. The data in the worksheets can be converted to a number of file formats. One of the most flexible file formats to accommodate questionnaire data is in the Microsoft Excel format.

Data that has been in the form of an excel file can then be processed automatically through the Data Analysis Tool provided by the user experience questionnaire (https://www.ueq-online.org/). Questionnaire entries were transformed into values to be processed.

The sequence of positive and negative terms for an item is randomly generated in the questionnaire. Per dimension half of the items start with positive and half with negative terms. The value +3 represents the most positive value and -3 the most negative value, as shown in Figure 2.

Figure 2. Questionnaire scale

### 3. RESULT AND DISCUSSION
The results and discussion section contains the characteristics of the respondent, the Go-Pay service in the Gojek application, the value and variance, the Pragmatic and Hedonic Quality, and the Benchmark Score.

### 3.1 Characteristics of Respondents
Total respondents involved in this research were 225 people. The respondents in this study consisted of a number of 2 (two) and 3 (three) level students who took the Human-Computer Interaction [33]–[36] and Research Methods [31], [37]–[39] course.

The respondents were dominated by male respondents by 57.33%, while female respondents totaled 42.46 percent. Then most of the respondents are included in generation Y (millennial) or millennial generation is 77.78 percent, then respondents from generation Z are 21.78 percent, and there are about 0.44 percent of respondents from generation X.

Table 3. Respondents' characteristics

|  | Characteristics | Percentage |
| --- | --- | --- |
| Gender | Male | 57.33 |
|  | Female | 42.67 |
| Generation | Generation X (gen Xers, busters) | 0.44 |
|  | Generation Y (gen Yers, millennials, nexters) | 77.78 |
|  | Generation Z | 21.78 |

### 3.2 Go-Pay Service on the Gojek Application

FinTech services on the Gojek Application are located on the online payment system, Go-Pay. The information position of the Go-Pay service is at the third section (Figure 3) which shows the active Go-Pay balance. Under the Go-Pay label there is a pay menu, promos, top up, and more.

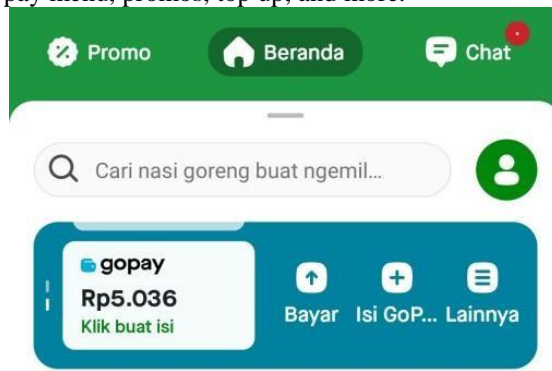

Figure 3. Display of go-pay in the gojek application

In the Go-Ride service payment scheme, the balance will be deducted automatically when a transaction between passenger and driver occurs. For payments to merchants that work with Gojek, there is a "Pay" button facility under Go-Pay, as shown in Figure 4. Click the button and navigate to the QR Code on the merchant.

One of the payment methods embedded in smartphones is technology "QR (Quick Response) codes" or "QR-Code" [40]. QR-Code or barcode to read an identity [26]. The Go-Pay service provides a number of ways to do transactions, namely: 1) Auto Debit, 2) Mobile Number, 3) Bank Account, and 4) Scan QR Code, as shown in Figure 4.

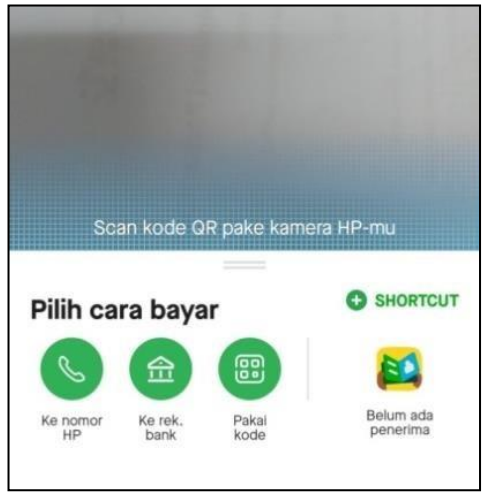
Figure 4. Go-pay payment options

Through the Regulation of Members of the Board of Governors "No: 21/18/PADG/2019 chapters 4 and 5", Bank Indonesia encourages the use of QR Code in conducting digital transactions. The term used is "Quick Response Code Indonesian Standard" hereinafter referred to QRIS. QRIS is expected to facilitate non-cash transactions of Indonesian society.

### 3.3 Mean and Variance
The mean and variance of each can be seen in Table 4. The mean range starts from 1.310 to 1.528. Values between -0.8 and 0.8 represent evaluations that are more or less neutral than the corresponding scale, values> 0.8 represent positive evaluations and values <-0.8 represent negative evaluations. The scale range is between -3 (very bad) and +3 (very good).

Table 4. Mean and variance

| No | Variable | Mean | Variance |
|---|---|---|---|
| 1 | Attractiveness | 1,389 | 1,62 |
| 2 | Perspicuity | 1,387 | 1,63 |
| 3 | Efficiency | 1,429 | 1,64 |
| 4 | Dependability | 1,248 | 1,59 |
| 5 | Stimulation | 1,417 | 1,58 |
| 6 | Novelty | 0,982 | 1,18 |

The mean statistical results describe the respondents' mean answers to the Go-Pay service. The highest mean score obtained by the variable "Efficiency" then followed by the variables "Stimulation", "Attractiveness", "Perspicuity", "Dependability", and the lowest variable "novelty". The statistical description of the mean, as shown in Figure 5.

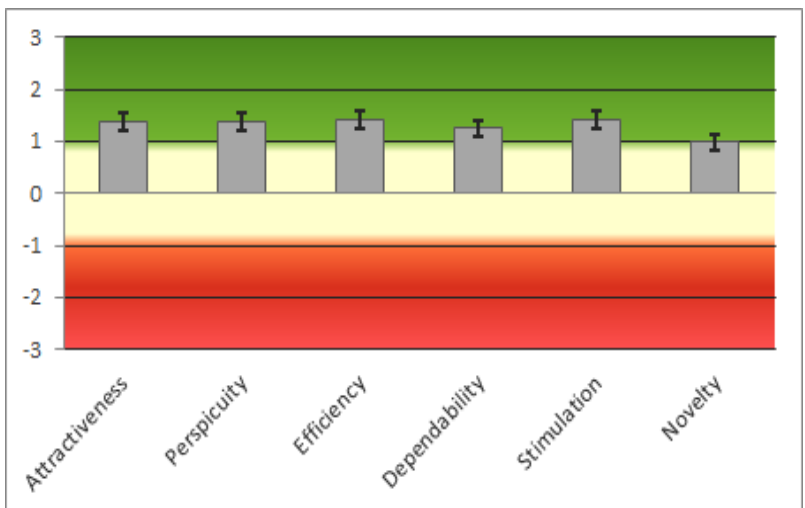

Figure 5. Mean score of go-pay services

The highest Go-Pay service variance score is obtained by the "Efficiency" variable, followed by the , "Perspicuity", "Attractiveness", "Dependability", "Stimulation", variables, and the lowest is the "Novelty" variable. The statistical description of the mean, as shown in Figure 6.

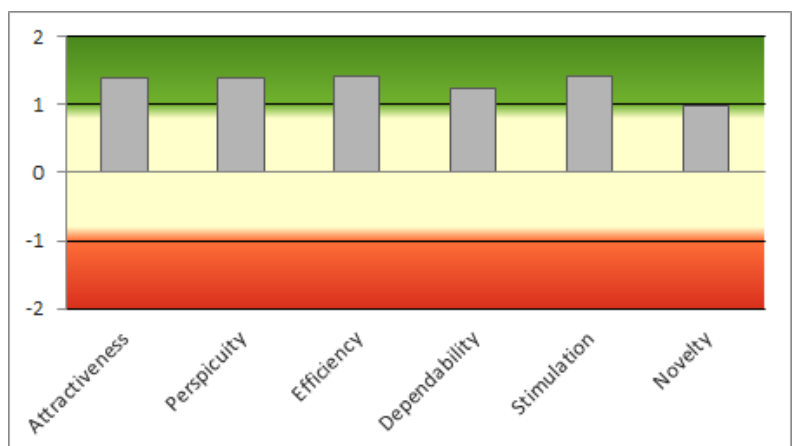

Figure 6. Variance score of go-pay services

### 3.4 Pragmatic and Hedonic Quality
The UEQ scale can be grouped into 2 (two), namely: 1) pragmatic quality and 2) hedonic quality (Table 5). Pragmatic quality consists of: a) Perspicuity, b) Efficiency, and c) Dependability). Whereas hedonic quality consists of: 1) Stimulation, and 2) Originality.

Table 5. Pragmatic and hedonic quality

| Category | Score |
|---|---|
| Attractiveness | 1,39 |
| Pragmatic Quality | 1,35 |
| Hedonic Quality | 1,20 |

Pragmatic quality describes aspects of quality related to tasks, hedonic quality, aspects of quality related to non-tasks. The results of the mean calculation of the three aspects of pragmatic and hedonic quality, as shown in Figure 7.

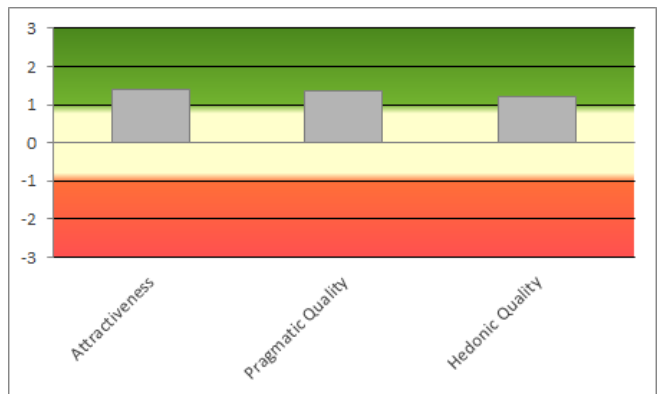

Figure 7. Pragmatic and hedonic quality score of go-pay services

### 3.5 Benchmark Score

The user experience questionnaire is also equipped with benchmark data. The results of the analysis are also complemented by benchmarks against a number of other products. The average scale measured is regulated in relation to the values available from the benchmark data set. This data set contains data from 20190 persons from 452 studies concerning different products (business software, web pages, web shops, social networks).

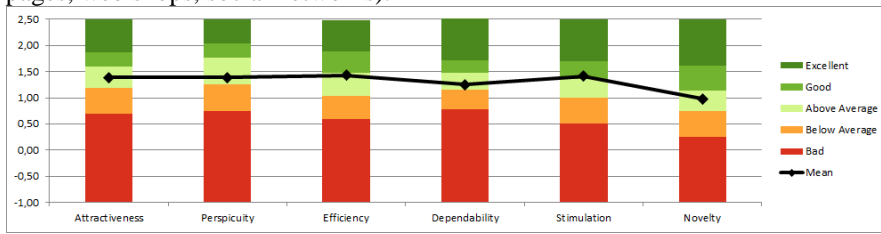

Figure 8. Benchmark score of go-pay services

Comparison of results for products evaluated with data in the benchmark allows conclusions about the relative quality of the product being evaluated compared to other products. The highest score sequence is achieved by variable 1) "Efficiency", followed by variables 2) "Stimulation", 3) "Attractiveness", 4) "Perspicuity", 5) Dependability, and 6) "Novelty".

Table 6. Benchmark comparison interpretation

| No | Variables | Mean | Comparison to benchmark | Interpretation |
|---|---|---|---|---|
| 1 | Attractiveness | 1,39 | Above average | 25% of results better, 50% of results worse |
| 2 | Perspicuity | 1,39 | Above Average | 25% of results better, 50% of results worse |
| 3 | Efficiency | 1,43 | Above Average | 25% of results better, 50% of results worse |
| 4 | Dependability | 1,25 | Above Average | 25% of results better, 50% of results worse |
| 5 | Stimulation | 1,42 | Good | 10% of results better, 75% of results worse |
| 6 | Novelty | 0,98 | Above Average | 25% of results better, 50% of results worse |

## 4. CONCLUSION

This paper filled the research gap in analysis FinTech application in Indonesia, especially related to user experience. The results of the study as a whole showed that all user experience variables for Go-Pay applications are above the average and good categories. But none of the variables reached the score in the excellent category. There is only one variable that gets a score in the "good" category, that is, the "stimulation" variable. Payment for Gojek services such as Go-Car and Go-Ride that can be automatically paid as long as the balance of Go-Pay contains makes the Go-Pay service considered to have a good level of efficiency. With big data analysis, Go-Pay services can read service trends that are commonly used by Gojek customers, so that it can facilitate the Gojek to provide stimulation in the form of promos that are tailored to the needs and habits of the customer. There are five variables that get a score in the "above average" category based on benchmark scores, namely: 1) Attractiveness, 2) Perspicuity, 3) Efficiency, 4) Dependability, and 5) novelty. The highest score sequence is achieved by variable 1) "Efficiency", followed by variables 2) "Stimulation", 3) "Attractiveness", 4) "Perspicuity", 5) Dependability, and 6) "Novelty". The novelties in the articles of this research include: 1) Research objects that have never been analyzed with previous user experience, 2) Respondents who are mostly millennial and z generations, 3) The results of the analysis are also equipped with benchmarks against a number of other products. For further research, the authors are interested in finding out the user experience of Grab users, and FinTech, OVO payment services. As well as making comparisons for the two FinTech applications.